\documentstyle[prd,preprint,aps,epsf]{revtex}

\begin{document}

\draft

\preprint{$\begin{array}{l}\\[-2cm]
           \mbox{\bf KEK--TH--441}\\[-3mm]
           \mbox{\bf KEK preprint 95--71}\\[-3mm]
           \mbox{\bf TU-482}\\[-3mm]
           \mbox{\bf TMUP-HEL-9504}\\[-3mm]
           \mbox{June~1995} \\[-3mm]
           \mbox{\bf H}\\[1cm]
           \end{array}$}
\title{Model-independent analysis of $\bbox{B}$-$\bbox{\bar B}$
       mixing\\
       and $\bbox{CP}$ violation in $\bbox{B}$ decays}
\author{Toru Goto$^1$, Noriaki Kitazawa$^2$,
        Yasuhiro Okada$^3$,
        and Minoru Tanaka$^3$}
\address{$^1$Department of Physics, Tohoku University,
         Sendai 980-77, Japan\\
         $^2$Department of Physics, Tokyo Metropolitan University,
         Tokyo 192-03, Japan\\
         $^3$Theory Group, KEK, Tsukuba, Ibaraki 305, Japan}
\maketitle
\begin{abstract}
We present a framework to analyze effects of
new physics beyond the standard model on $B$-$\bar B$
mixing and $CP$ violation in $B$ decays in a model-independent
manner. Assuming that tree level decay amplitudes are dominated
by the standard model ones, new physics
contribution to the $B$-$\bar B$ mixing can be extracted from
several measurements at $B$ factories. Using this framework,
we show the present
constraint on new physics contribution to
the $B$-$\bar B$ mixing, and illustrate constraints
expected to be given by future experiments at $B$ factories.
We also point out a possibility that $CP$ asymmetries in
$B\rightarrow\psi K_S$, $B\rightarrow\pi\pi$,
and $B\rightarrow DK$ modes look consistent with
the standard model, even if a large new physics
contribution is present in the $B$-$\bar B$ mixing.
\end{abstract}

Physics of $B$ meson provides several tests of
the standard model and could give insights into new physics
beyond it. Especially, in the standard model, the test of
unitarity of the Cabibbo-Kobayashi-Maskawa (CKM) matrix\cite{KM}
is most important. As shown in Fig.~\ref{UT},
the unitarity of the CKM matrix is graphically expressed
by a triangle.
In the standard model, the lengths of the sides
are related to several decay rates and/or the magnitude of
the $B$-$\bar B$ mixing, while the angles are related to
several $CP$ asymmetries. When these quantities are
measured at future $B$ factories, we will be able to test
the unitarity of the CKM matrix by looking whether
the triangle is closed or not.

Quantitatively, the analysis will be done along
the $\chi^2$ test: All the observables such as decay rates and
$CP$ asymmetries are represented by the standard model parameters
like the CKM parameters and some hadronic
parameters like the $B$ meson decay constant $f_B$.
The $\chi^2$ is calculated with
experimental results of these quantities and knowledge of the
hadronic parameters as a function of the standard model parameters.
Then, this $\chi^2$ is minimized by varying the standard model
parameters, and we can see
whether the standard model is consistent or not
depending on the obtained $\chi^2_{\mbox{min}}$.

Although it is quite straightforward
and powerful, this method has some defects.
Even if a larger value of $\chi^2_{\mbox{min}}$ is
found, this method itself does not tell us anything other than
that the standard model is doubtful. No quantitative
information about new physics can be obtained.
Moreover, even if the standard model seems to be consistent,
there may be new physics which evades the above $\chi^2$
test. From these points of view,
it is desirable to introduce
some new physics effects into the analysis and to see
how they are restricted by experiments.

For this purpose, two approaches are possible:
One is a model-dependent approach, in which we
specify a model of new physics and analyze all data in it.
Although each model needs a specific analysis,
this approach has a predictive power if the model does not have
too many parameters. Another approach is
a model-independent approach, in which we do not
introduce any model. The data are analyzed based on
rather general principles or assumptions. If we can
parametrize effects of new physics by a few model-independent
quantities, this approach is quite suitable to
select appropriate new physics among many possibilities
based on experimental information.
We pursue the latter approach in this paper.

Weak interactions of $B$ meson are described by
$|\Delta B|=1$ and $|\Delta B|=2$ amplitudes,
{\em i.e.\/} $B$ decay and $B$-$\bar B$ mixing
amplitudes respectively.
In the standard model, the $|\Delta B|=1$ processes
occur through the tree and penguin diagrams
at the quark level.
Effects of new physics tend to appear in a penguin
diagram because it is a loop diagram.
In fact, in some classes of new physics like SUSY, tree-level
$B$ decay amplitudes are hardly affected by new physics.
On the other hand, it is difficult to exclude
effects of new physics in the penguin diagrams.

In order for the model-independent approach,
we classify $B$ decay processes into two classes.
Although we are considering general cases,
it is useful to introduce class I and class II
based on the properties of the quark-level amplitudes
in the standard model.
The class I processes do not have the penguin part in
its quark-level amplitude, and the class II processes do.
The processes of $b\rightarrow q_d q_u\bar q_u'$ type,
where $q_d$ represents a down-type quark, and
$q_u$ and $q_u'$ denote different up-type quarks, are
class I processes. The ordinary semileptonic decays are
also regarded as class I processes because their amplitudes
have no penguin part.
While, the processes of $b\rightarrow q_d q\bar q$ type,
where $q$ denotes a generic type of quark, are class II processes.
If $q$ is an up-type quark, $b\rightarrow q_d q\bar q$ consists of
the penguin part and the tree part.
If $q$ is a down-type quark, $b\rightarrow q_d q\bar q$
contains only the penguin part%
\footnote{The processes $b\rightarrow dd\bar s$ and
          $b\rightarrow ss\bar d$ do not appear in the
          standard model within the lowest order of
          the weak interaction. Even if these processes are
          caused by new physics, they do not affect
          the following discussions.}.

In the following, we assume that the class I
processes are described by the standard W-exchange diagrams.
In addition, we assume that the unitarity of the CKM matrix
is saturated by the first three generations. This assumption
is necessary to determine the quark-mixing parameters
from class I processes and
to apply them to evaluate the standard model contribution
to the $B$-$\bar B$ mixing as is described below.

New physics can also contribute to the $B$-$\bar B$ mixing
amplitude $M_{12}$. We can write it as
\begin{equation}
M_{12}=|M_{12}|e^{i\phi_M}=M_{12}^{\rm SM}+M_{12}^{\rm NEW},
\label{M12}
\end{equation}
where $M_{12}^{\rm SM}$ is the standard model contribution
and $M_{12}^{\rm NEW}$ represents contributions from new physics.
$M_{12}^{\rm NEW}$ is the model-independent parameter which
describes effects of new physics in the $B$-$\bar B$ mixing.
Note that Eq.~(\ref{M12}) is enough to determine a phase
convention for $M_{12}^{\rm NEW}$ once the phase convention
in evaluating $M_{12}^{\rm SM}$ is fixed, because the relative
phase between $M_{12}^{\rm SM}$ and $M_{12}^{\rm NEW}$ is
a physically meaningful quantity. We use the following
expression of $M_{12}^{\rm SM}$:
\begin{equation}
M_{12}^{\rm SM}=\frac{G_F^2}{12\pi^2}m_W^2m_Bf_B^2B_B\eta_B
                (V_{td}^*V_{tb})^2 S(m_t^2/m_W^2),
\label{M12SM}
\end{equation}
where
\begin{equation}
(V_{td}^*V_{tb})^2=\lambda^6A^2(1-\rho+i\eta)^2,
\end{equation}
in the Wolfenstein parametrization\cite{W}, and
\begin{equation}
S(x)=x\left[\frac{1}{4}+\frac{9}{4(1-x)}-\frac{3}{2(1-x)^2}\right]-
     \frac{3}{2}\frac{x^3}{(1-x)^3}\ln{x},
\end{equation}
is the Inami-Lim function\cite{IL}.

In the following, we see how $M_{12}^{\rm NEW}$ is constrained
by the present experiments, and how the constraint will be
improved by future experiments. For this purpose, we have to
determine the Wolfenstein parameters $A$, $\rho$, and $\eta$
simultaneously with $M_{12}^{\rm NEW}$.
The ordinary analysis in which the standard model
is assumed cannot be applied if we consider effects of new physics.
However, according to the above two assumptions,
{\em i.e.\/} the dominance of the standard model contributions
in the class I processes and the three-generation unitarity,
we can extract information
about the Wolfenstein parameters without being bothered by
the unknown new physics in the class II processes
which is not parametrized in our analysis.

First, let us consider the semileptonic decay of the $B$ meson,
$\bar B\rightarrow X_c\ell\bar\nu$, which
is one of the class I processes
and free from new physics.
We can determine the CKM matrix element $|V_{cb}|=\lambda^2 A$
from its width.
Secondly, the charmless semileptonic decay of the $B$ meson,
$\bar B\rightarrow X_u\ell\bar\nu$
is also considered to be free from new physics, and we can obtain
a constraint on $|V_{ub}/V_{cb}|=\lambda\sqrt{\rho^2+\eta^2}$
from its rate.
These exhaust the presently available constraints
which are not affected by new physics in our framework.

The remaining constraint relevant to our analysis is
the one given by the observation of $B$-$\bar B$ mixing.
{}From the experiments, we obtain a constraint on the absolute
value of the $B$-$\bar B$ mixing amplitude,
$|M_{12}|=|M_{12}^{\rm SM}+M_{12}^{\rm NEW}|$.
Since the possible range of $M_{12}^{\rm SM}$ is limited
by the above mentioned constraints on the Wolfenstein parameters
and the knowledge of the top quark mass and the hadronic
matrix element,
we can obtain an allowed region of $M_{12}^{\rm NEW}$.
We do not include the information from the $K$-$\bar K$ mixing
into our analysis because we need to introduce another
model-independent parameter, $M_{12}^{\rm NEW}$ for the $K$-$\bar K$
mixing.

Here, we present a result of the analysis which follows the above
strategy. We summarize the inputs in Table \ref{PE}.
$\lambda$ is determined by the semileptonic kaon and
hyperon decays\cite{M}, which are free from new physics in our
framework. $A$ and $\sqrt{\rho^2+\eta^2}$ are constrained by
the semileptonic $B$ decays as mentioned above\cite{S}.
$|M_{12}|$ is obtained from the measurement of
the $B^0$-$\bar B^0$ mass difference $\Delta m$\cite{BH}.
The top quark mass $m_t$ is given
by the direct observations\cite{TOP}.
$f_B\sqrt{B_B\eta_B}$ is estimated by several theoretical methods.
We take the value used in Ref.~\cite{R}.
We omitted experimental errors and/or
theoretical uncertainties of $\lambda$, $A$, $m_t$,
and $f_B\sqrt{B_B\eta_B}$ for simplicity.

Fig.~\ref{M12PLANE} shows the experimentally allowed region and
the standard model prediction for $M_{12}$ in the complex
$M_{12}$ plane. The circles whose centers are at the origin
show the direct experimental constraint on $|M_{12}|$
in Table \ref{PE}, and the distorted circles show
the standard model ``prediction'' which is calculated by
Eq.~(\ref{M12SM}) with
the inputs in Table \ref{PE}. A vector from a point in the region
of the standard model prediction to a point in the experimentally
allowed region corresponds to a possible complex value of
$M_{12}^{\rm NEW}$. From this figure we can see that the value of
$M_{12}^{\rm NEW}$ is certainly limited. By integrating out
$\rho$ and $\eta$ in the maximum likelihood method,
we can obtain an allowed region of $M_{12}^{\rm NEW}$. The result is
shown in Fig.~\ref{M12NEW}(a), where we show the line that
$\chi^2=\chi^2_{\rm min}+4.61$, which is usually regarded as
90\% confidence level. This figure tells us that
the contribution of $M_{12}^{\rm NEW}$ to the total
$B$-$\bar B$ mixing amplitude $M_{12}$ can be comparable with
or larger than that of $M_{12}^{\rm SM}$, besides that
the standard model is completely consistent.
Also note that a positive value of ${\rm Re}M_{12}^{\rm NEW}$
tends to be disfavored.

Now, let us discuss effects of new physics on $CP$ asymmetries
in $B$ decays which will be measured in future experiments at
$B$ factories. We start with the class I process
$B\rightarrow DX_s$, for which we can use the standard model
decay amplitudes. For the charged $B$ meson mode
$B^\pm\rightarrow DX_s^\pm$, since this decay mode
is free from new physics provided that no significant
$CP$ violation in $D^0$-$\bar D^0$ mixing is caused by new physics,
the method by Gronau and Wyler\cite{GW} can be applied to extract
information of the angle $\phi_3$ defined in Fig.~\ref{UT}.
We can uniquely determine
$\sin^2\phi_3$ using two strange states which make
different phase shifts from each other,
{\em e.g.\/} $K$ and $K^*$, for $X_s$.
For the neutral $B$ meson mode
$B^0(\bar B^0)\rightarrow DK_S$\cite{GL1},
the situation is a little different
owing to the new physics contribution to the $B$-$\bar B$
mixing, $M_{12}^{\rm NEW}$. From the time-dependent decay rate,
however, we can uniquely determine $\sin{\phi_M}$,
$\sin(\phi_M+2\phi_3)$, and $\sin^2\phi_3$%
\footnote{In the standard model,
          which is considered in Ref.~\cite{GL1},
          $\phi_M$ defined in Eq.~(\ref{M12}) corresponds to
          $2\phi_1$ in our phase convention.}.

For $B\rightarrow \pi\pi$ process, the decay amplitude can be
affected by new physics because it is a class II process.
However, we can extract the $\Delta I=3/2$ part of the amplitude
by the isospin analysis\cite{GL2}. Since  the $\Delta I=3/2$ part
of the amplitude does not contain the penguin contribution,
it can be regarded as a class I process.
Although this analysis depends on
the assumption that the penguin part of the amplitude is
$\Delta I=1/2$, it seems to be plausible even in the presence
of new physics. In other words, we do not expect that
the electromagnetic penguin, the electroweak penguin,
and the box diagram play important roles in this process
even in the presence of new physics
as in the standard model\cite{DH}.
In such a case, we can extract $\sin(\phi_M+2\phi_3)$
with fourfold ambiguity. If the penguin contribution turns out to
be small, this fourfold ambiguity disappears. We can also resolve
it by comparing with the $\sin(\phi_M+2\phi_3)$ obtained from
$B^0(\bar B^0)\rightarrow DK_S$.

The last process discussed here is
$B^0(\bar B^0)\rightarrow\psi K_S$, which is classified
as a class II process.
In the standard model, the penguin contribution in this
process does not cause any problem, because the weak
phases of the tree and the penguin contributions are the same.
However, they could differ from each other in the presence of
new physics, so that we cannot extract information about
$\phi_M$ from this process in general. Nevertheless,
it is desirable to include this mode in our analysis because
it is expected to be precisely measured.
In the following, we assume that
there is no significant penguin contribution which has a different
$CP$-violating phase from that of the standard model
in this process. With this assumption,
we can uniquely determine $\sin\phi_M$ from the time-dependent
decay rate. Note that the above assumption can be checked by
comparing the obtained $\sin\phi_M$ in this process with
that obtained in $B^0(\bar B^0)\rightarrow DK_S$ process.
Also, it can be tested by looking for
the $CP$-violating rate differences
(the direct $CP$ violation) which may be seen in
$B^\pm\rightarrow \psi K^\pm$ and $\cos(\Delta m\,t)$ term
in the time-dependent rate of $B^0(\bar B^0)\rightarrow\psi K_S$
decay depending on the relevant phase shifts.
Table \ref{INFO} summarizes the above arguments on
the several modes of studying $CP$ violation.

Now, let us illustrate how measurements of $CP$ asymmetries in
these modes at $B$ factories constrain the model-independent
parameter $M_{12}^{\rm NEW}$. The inputs for this illustration
in addition to those in Table \ref{PE} are given in
Table \ref{FE}. The central values of these inputs are
calculated by putting $M_{12}^{\rm NEW}=0$ and
$(\rho,\eta)=(0.23,0.29)$ which is a typical point allowed in
the standard model analysis\cite{R}.
The errors in this table
are taken from Ref.~\cite{LOI}.

At an earlier stage of a $B$ factory run,
we will observe only the $CP$ violation in
$B^0(\bar B^0)\rightarrow \psi K_S$
mode. In Fig.~\ref{M12NEW}(b), we show the expected constraint
on $M_{12}^{\rm NEW}$ from this mode which gives
the constraint on $\sin\phi_M$ and the information given
in Table \ref{PE}. The expected constraint is not much stronger
than that in Fig.~\ref{M12NEW}(a), because $\phi_3$ still remains
free. In Fig.~\ref{M12NEW}(c), we show the expected constraint
by adding the information from $B^\pm\rightarrow DX_s^\pm$
{\em i.e.\/} the constraints on $\sin^2\phi_3$.
In this case, we have eight solutions
for $M_{12}^{\rm NEW}$ owing to twofold ambiguity in $\phi_M$
determination from $\sin\phi_M$ and fourfold ambiguity in
$\phi_3$ determination from $\sin^2\phi_3$.
These solutions are indicated
by the dots in Fig.~\ref{M12NEW}(c).
The rather larger allowed regions in
Fig.~\ref{M12NEW}(c) consist of those surrounding
these eight solutions.
Fig.~\ref{M12NEW}(d) shows the expected constraint from
all the inputs
given in Table \ref{PE} and \ref{FE}. We have four solutions
even in this case.
If allowed regions obtained with real data
do not contain the origin ($M_{12}^{\rm NEW}=0$), we can
not only conclude that the standard model is excluded but also
determine the new physics contribution quantitatively.

An important observation from this illustration
is that we cannot exclude the possibility of significantly large
$M_{12}^{\rm NEW}$ even if all the measurements
considered here seem to be
consistent with the standard model.

We have neglected the uncertainties of $A$, $m_t$, and
$f_B\sqrt{B_B\eta_B}$ in the above analysis. The allowed
regions in Figs.~\ref{M12NEW}(a)--(d) are changed
if we vary these values. However, we expect that these
uncertainties will be reduced enough by future
experimental and theoretical developments.

In conclusion, we presented a framework of analysing the
$B$-$\bar B$ mixing and the $CP$ violations in $B$ decays
in a model-independent manner. We introduced the
model-independent parameter $M_{12}^{\rm NEW}$ and showed
the constraint on it from the presently available experimental data.
We also illustrated how $M_{12}^{\rm NEW}$ would be constrained
by the future experiments of $CP$ violation at $B$ factories.
We found that there remain some non-trivial solutions
of $M_{12}^{\rm NEW}$ even for the inputs corresponding to
the standard model case ($M_{12}^{\rm NEW}=0$).
This means that we cannot exclude the
possibility of significant contribution to
the $B$-$\bar B$ mixing from new physics even if
$CP$ asymmetries in all the modes of $B\rightarrow\psi K_S$,
$B\rightarrow DX_s$, and $B\rightarrow \pi\pi$
seem to be consistent with the standard model.

The authors would like to thank C.~S.~Lim for useful discussions.

\newpage
\begin{table}
\begin{tabular}{cc}
$\lambda$ & 0.220\cite{M} \\
\hline
$|V_{cb}|=\lambda^2 A$ & 0.038\cite{S} \\
\hline
$m_t$ & 174 GeV\cite{TOP} \\
\hline
$f_B\sqrt{B_B\eta_B}$ & 165 MeV\cite{R} \\
\hline
$|V_{ub}/V_{cb}|=\lambda\sqrt{\rho^2+\eta^2}$ &
                                      $0.08\pm 0.02$\cite{S} \\
\hline
$\Delta m=2|M_{12}|$ & $0.462\pm 0.026\;{\rm ps}^{-1}$\cite{BH} \\
\end{tabular}
\caption{The inputs corresponding to the present
         experimental and theoretical knowledge.}
\label{PE}
\end{table}

\begin{table}
\begin{tabular}{ll}
$B^\pm\rightarrow DX_s^\pm$ & $\sin^2\phi_3$ \\
\hline
$B^0(\bar B^0)\rightarrow DK_S$ &
        $\sin\phi_M,\,\sin(\phi_M+2\phi_3),\,\sin^2\phi_3$ \\
\hline
$B\rightarrow\pi\pi$ & $\sin(\phi_M+2\phi_3)$ (fourfold in general)\\
\hline
$B^0(\bar B^0)\rightarrow \psi K_s$ &
        $\sin\phi_M$ (assuming no penguin which
                      has a non-standard phase)\\
\end{tabular}
\caption{Quantities obtained from several processes of $CP$
         violation search.}
\label{INFO}
\end{table}

\begin{table}
\begin{tabular}{cc}
$\sin\phi_M$ & $0.66\pm 0.08$ \\
\hline
$\sin^2\phi_3$ & $0.62\pm 0.25$ \\
\hline
$\sin(\phi_M+2\phi_3)$ & $0.57\pm 0.17$ \\
\end{tabular}
\caption{The inputs used for the illustration of the expected
         constraints from $CP$ violation experiments
         at $B$ factories.}
\label{FE}
\end{table}


\begin{figure}
\caption{The unitarity triangle.}
\label{UT}
\end{figure}

\begin{figure}
\caption{The experimentally allowed region and the standard model
         prediction for $M_{12}$. The circles whose centers
         are at the origin show the direct experimental constraint,
         and the distorted circles shows the standard model
         prediction. The dashed lines show the standard model
         predictions for several fixed $\phi_3$ values. (The
         angle $\phi_3$ is defined in Fig.~\protect\ref{UT}.)}
\label{M12PLANE}
\end{figure}

\begin{figure}
\caption{The allowed regions of $M_{12}^{\rm NEW}$.
         The lines that $\chi^2=\chi^2_{\rm min}+4.61$ are shown.
         (a) The presently allowed region  corresponding to
             the inputs in Table \protect\ref{PE}.
         (b) The expected constraint from $B\rightarrow\psi K_S$
             in addition to the inputs in Table \protect\ref{PE}.
         (c) The expected constraint from $B\rightarrow\psi K_S$
             and $B^\pm\rightarrow DX_s^\pm$ in addition to
             the inputs in Table \protect\ref{PE}.
             The dots represent the solutions obtained from
             the central values of the inputs.
         (d) The expected constraint from
             all the inputs listed in Table \protect\ref{PE}
             and \protect\ref{FE}. The dots mean the same as (c).}
\label{M12NEW}
\end{figure}

\begin{center}
\newpage
\vspace*{15ex}
\epsfbox{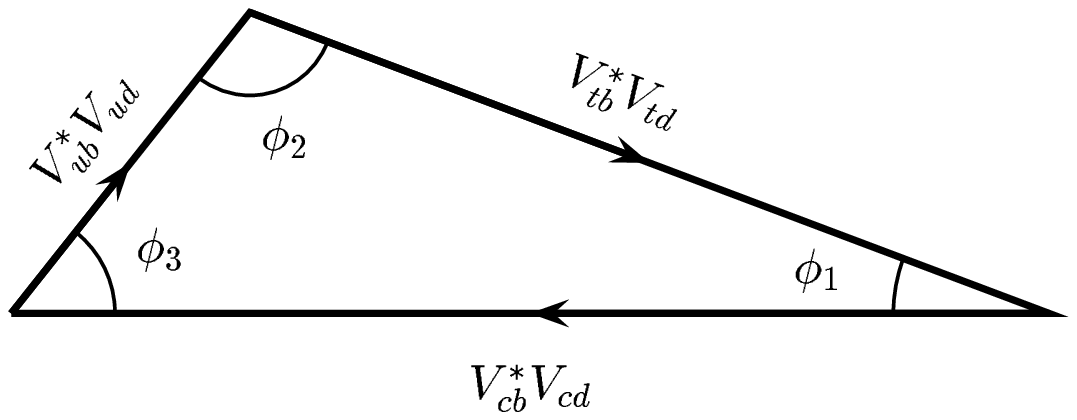}
\LARGE{Fig.~\ref{UT}}
\newpage
\epsfbox{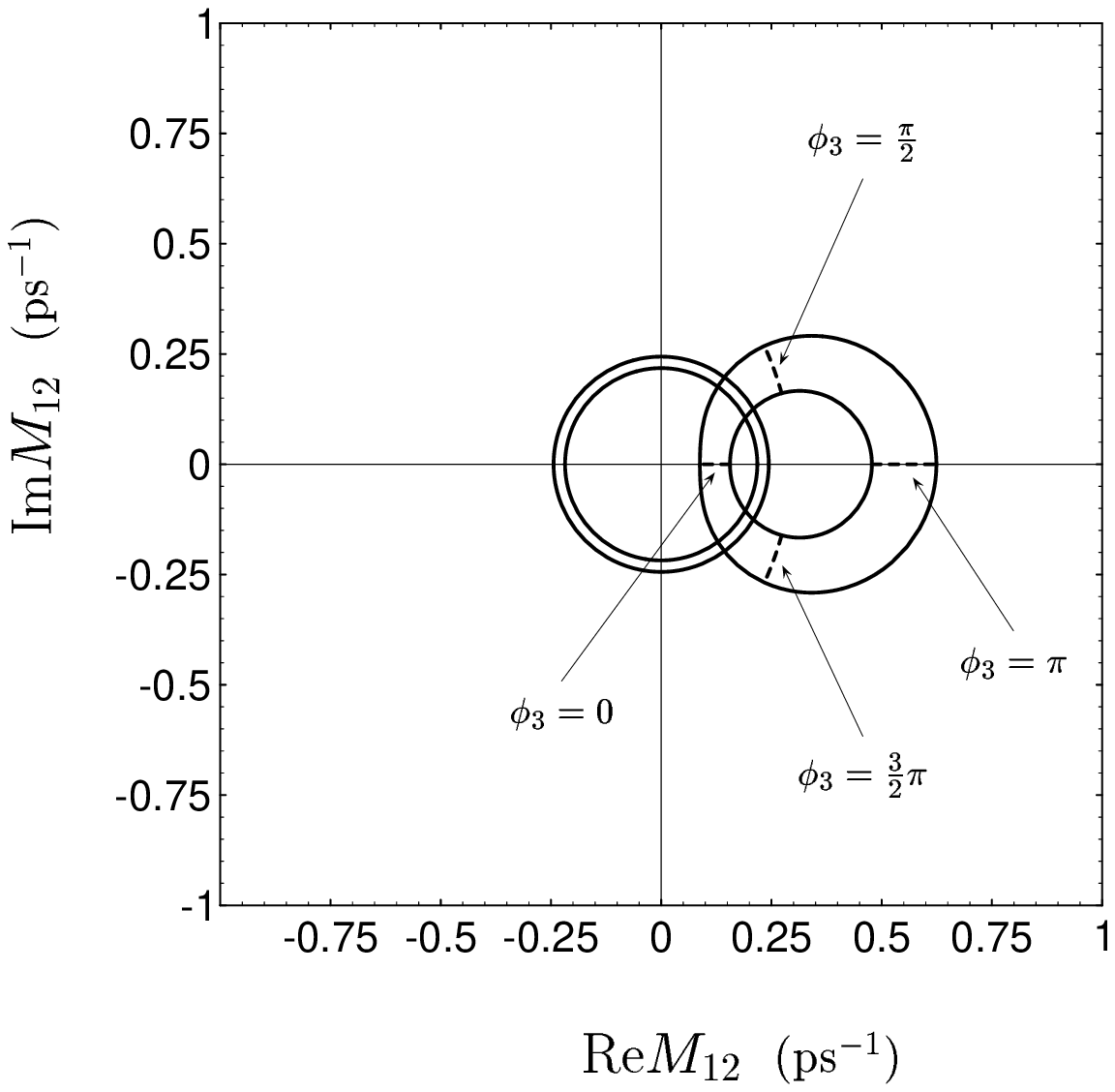}
\LARGE{Fig.~\ref{M12PLANE}}
\newpage
\epsfbox{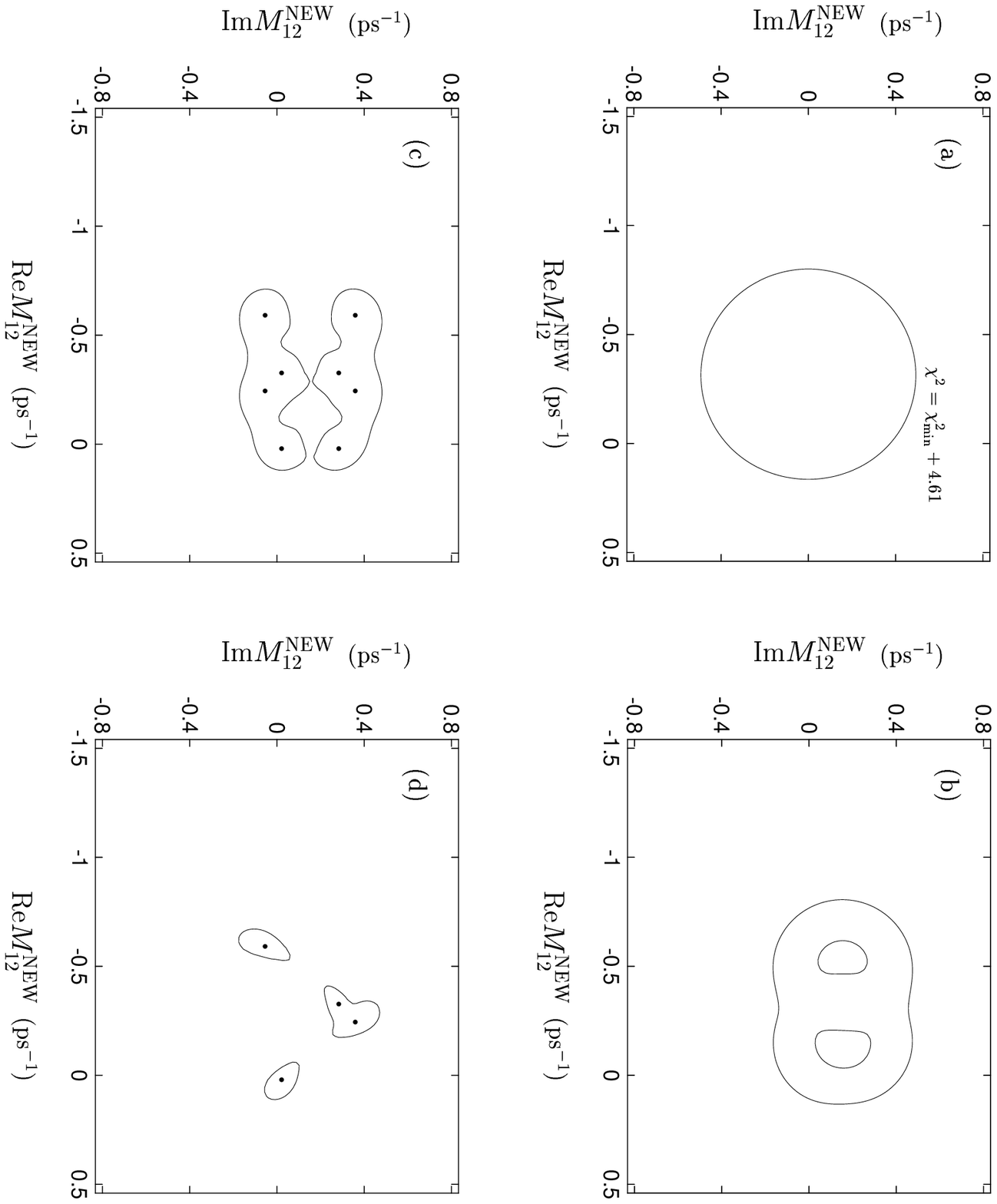}
\LARGE{Fig.~\ref{M12NEW}}
\end{center}

\end{document}